\documentclass[prl,reprint,twocolumn,showpacs,amsmath,amssymb,aps,superscriptaddress]{revtex4}

\usepackage{graphicx,bm,times,color}

\newcommand{\op}[1]{\hat{#1}}

\begin{document}

\title{Assessing the Polarization of a  Quantum Field from Stokes Fluctuations} 

\author{A.~B.~Klimov} 
\affiliation{Departamento de F\'{\i}sica,
Universidad de Guadalajara, 
44420~Guadalajara, Jalisco, Mexico}

\author{G.~Bj\"{o}rk} 
\affiliation{School of Communication and Information Technology, 
Royal Institute of Technology (KTH), Electrum 229, 
SE-164 40 Kista, Sweden}

\author{J.~S\"{o}derholm} 
\affiliation{School of Communication and Information Technology, 
Royal Institute of Technology (KTH), Electrum 229, 
SE-164 40 Kista, Sweden}
\affiliation{Max-Planck-Institut f\"ur die Physik des Lichts, 
G\"{u}nther-Scharowsky-Stra{\ss}e 1, Bau 24, 
91058 Erlangen,  Germany}
\affiliation{Universit\"{a}t Erlangen-N\"{u}rnberg,
Staudtstra{\ss}e 7/B2, 91058 Erlangen, Germany}

\author{L. S. Madsen}
\affiliation{Department of Physics, 
Technical University of Denmark, Building 309, 
2800 Kongens Lyngby, Denmark}

\author{M. Lassen}
\affiliation{Department of Physics, 
Technical University of Denmark, Building 309, 
2800 Kongens Lyngby, Denmark}

\author{U. L. Andersen}
\affiliation{Department of Physics, 
Technical University of Denmark, Building 309, 
2800 Kongens Lyngby, Denmark}

\author{J. Heersink}
\affiliation{Max-Planck-Institut f\"ur die Physik des Lichts,
 G\"{u}nther-Scharowsky-Stra{\ss}e 1, Bau 24, 
91058 Erlangen,  Germany}
\affiliation{Universit\"{a}t Erlangen-N\"{u}rnberg,
Staudtstra{\ss}e 7/B2, 91058 Erlangen, Germany}

\author{R. Dong}
\affiliation{Max-Planck-Institut f\"ur die Physik des Lichts,
 G\"{u}nther-Scharowsky-Stra{\ss}e 1, Bau 24, 
91058 Erlangen,  Germany}
\affiliation{Universit\"{a}t Erlangen-N\"{u}rnberg,
Staudtstra{\ss}e 7/B2, 91058 Erlangen, Germany}

\author{Ch.~Marquardt}
\affiliation{Max-Planck-Institut f\"ur die Physik des Lichts,
 G\"{u}nther-Scharowsky-Stra{\ss}e 1, Bau 24, 
91058 Erlangen,  Germany}
\affiliation{Universit\"{a}t Erlangen-N\"{u}rnberg,
Staudtstra{\ss}e 7/B2, 91058 Erlangen, Germany}
 
\author{G.~Leuchs}
\affiliation{Max-Planck-Institut f\"ur die Physik des Lichts, 
G\"{u}nther-Scharowsky-Stra{\ss}e 1, Bau 24, 
91058 Erlangen,  Germany}
\affiliation{Universit\"{a}t Erlangen-N\"{u}rnberg,
Staudtstra{\ss}e 7/B2, 91058 Erlangen, Germany}

\author{L.~L.~S\'{a}nchez-Soto} 
\affiliation{Max-Planck-Institut f\"ur die Physik des Lichts, 
G\"{u}nther-Scharowsky-Stra{\ss}e 1, Bau 24, 
91058 Erlangen,  Germany}
\affiliation{Universit\"{a}t Erlangen-N\"{u}rnberg,
Staudtstra{\ss}e 7/B2, 91058 Erlangen, Germany}

\begin{abstract}
  We propose an operational degree of polarization in terms of the
  variance of the projected Stokes vector minimized over all the
  directions of the Poincar\'e sphere. We examine the properties of
  this degree and show that some problems associated with the standard
  definition are avoided. The new degree of polarization is
  experimentally determined using two examples: a bright squeezed
  state and a quadrature squeezed vacuum.
\end{abstract}


\date{\today}
  
\maketitle

\textit{Introduction.---}
Polarization is a fundamental property of light that has received a
lot of attention over the years~\cite{Brosseau:1998lr}. Nowadays, the
topic is witnessing a revival in interest because of the fast
developments, both on applications and on more fundamental aspects. As
polarization is a robust characteristic, relatively simple to
manipulate without inducing more than marginal losses, it is not
surprising that many experiments at the forefront of quantum optics
involve this observable~\cite{Mandel:1995qy}.

In classical optics, polarization can be elegantly visualized by using
the Poincar\'e sphere and is determined by the Stokes parameters.
These are directly measurable quantities that can be straightforwardly
extended to the quantum domain, where they become
operators~\cite{Luis:2000ys}.

The classical degree of polarization is simply the length of the
Stokes vector. While this provides a very intuitive picture, for many
complex fields it has also serious drawbacks. Indeed, this classical
quantity does not distinguish between states having remarkably
different polarization properties~\cite{Tsegaye:2000mz}. In
particular, it can be zero for light that cannot be regarded as
unpolarized, giving rise to the so-called ``hidden
polarization''~\cite{Klyshko:1992wd}.

These flaws have prompted some novel generalizations of the degree of
polarization~\cite{Luis:2002ul,Legre:2003qf,Saastamoinen:2004lq,
  Ellis:2005cr,Luis:2005qf,Sehat:2005wd}. A notion that has been
gaining support in the quantum optics community is to apply a properly
chosen distance~\cite{Klimov:2005kl} (entropy can be regarded as a
special case~\cite{Refregier:2005uq}).  This has the potential
advantage of circumventing most of the aforementioned difficulties,
while making close contact with other measures introduced to quantify
quantum resources~\cite{Fuchs:1996qy}.

There is, however, a problem with this approach: these distances can
be computed (not measured) only after a complete knowledge of the
state, which in practice implies a full quantum tomography. In other
words, while offering very good properties, distance measures do not
have a clear operational meaning.

We adhere to the view that the Stokes variables constitute a natural
tool in appraising polarization properties, so they should be the
basic building blocks for any practical degree of polarization. One
can expect that the problems arising with the classical degree
are due to its definition in terms exclusively of first-order moments
of the Stokes variables.  This may be sufficient for most classical
problems, but for quantum fields higher-order correlations are
crucial.

Our goal in this Letter is to provide such a characterization. We
learn from coherence theory that a full description of interference
phenomena may involve a hierarchy of degrees.  In this vein, we go
beyond the first-order description and look for a second-order degree
as the minimum Stokes variance over all directions of the Poincar\'e
sphere.  This simple proposal will prove very satisfactory when facing
the complications known in this field.  We also present a couple of
experimental examples confirming the feasibility of our scheme.

\textit{Polarization structure of quantum fields.---} 
We begin by briefly recalling some background material. We assume a
two-mode quantum field that is described by two complex amplitudes,
$\op{a}_{H}$ and $\op{a}_{V}$, where the subscripts $H$ and $V$
indicate horizontally and vertically polarization modes, respectively.
The commutation relations of these operators are standard: $[\op{a}_j,
\op{a}_k^\dagger ] = \delta_{jk}$, with $j, k \in \{H, V \}$. The
analysis is greatly simplified if we use the Stokes operators
\begin{eqnarray}
  \label{Stokop}
  & \op{S}_{x} =   \op{a}_{H}  \op{a}^\dagger_{V} + 
  \op{a}^\dagger_{H} \op{a}_{V}  \, ,
  \qquad
  \op{S}_{y} =  i ( \op{a}_{H} \op{a}^\dagger_{V} -
  \op{a}^\dagger_{H} \op{a}_{V} ) \, ,  & \nonumber \\
  & & \\
  & \op{S}_{z} =   \op{a}^\dagger_{H} \op{a}_{H} -
  \op{a}^\dagger_{V} \op{a}_{V}  \, , \nonumber &
\end{eqnarray}
together with the total photon number $\op{S}_{0} = \op{N} =
\op{a}^\dagger_{H} \op{a}_{H} + \op{a}^\dagger_{V} \op{a}_{V}$. The
average values of these operators are precisely the classical Stokes
parameters.  One immediately finds that the components of the Stokes
vector $\op{\mathbf{S}} = (\op{S}_{x}, \op{S}_{y}, \op{S}_{z})^{t}$
(where $t$ denotes the transpose) satisfy the commutation relations
distinctive of the su(2) algebra: $[\op{S}_{x}, \op{S}_{y}] = i 2
\op{S}_{z}$ and cyclic permutations. This noncommutability precludes
their simultaneous precise measurement, which is expressed by the
uncertainty relation
\begin{equation}
  \label{ur}
  (\Delta  \mathbf{S} )^{2} = 
  (\Delta S_{x} )^2 + (\Delta S_{y} )^2 + (\Delta S_{z} )^2  
  \geq  2  \langle \op{S}_{0} \rangle \, ,
\end{equation}  
with $(\Delta S_{k})^{2}$ ($k= x, y, z$) being the corresponding
variances.  In addition, $[\op{S}_{0}, \op{\mathbf{S}}] = 0$, so we
can treat each subspace with a fixed number of photons $N$
separately. This can be emphasized if instead of the Fock basis for
both polarization modes, $|n \rangle_{H} | m \rangle_{V}$ ($n, m = 0,
\ldots, \infty$), we employ the relabeling $| N, k \rangle = | k
\rangle_{H} |N - k \rangle_{V}$ ($k = 0, 1, \ldots, N$).  In this way,
for each fixed $N$, these states span an SU(2) invariant subspace of
dimension $N + 1$.

The standard definition of the degree of polarization for a quantum
state $\op{\varrho}$ is
\begin{equation}
  \label{eq:Psc}
  \mathbb{P}_{1} (\op{\varrho})  =
  \frac{| \langle \op{\mathbf{S}} \rangle |}
  {\langle \op{S}_{0} \rangle} =
  \frac{\sqrt{\langle \op{S}_{x} \rangle^2
      + \langle \op{S}_{y} \rangle^2 +
      \langle \op{S}_{z} \rangle^2}}
  {\langle \op{S}_{0} \rangle} \, ,
\end{equation}
where the subscript 1 stresses here that it involves first-order
moments of the Stokes variables.  We note that for any single-mode
state of the form $ | \Psi \rangle_{H} | 0 \rangle_{V}$, we get
$\mathbb{P}_{1} = 1$, which seems unphysical for a variety of
reasons. In particular, when $| \Psi \rangle_{H} \rightarrow | 0
\rangle_{H}$, we have $\mathbb{P}_{1} = 1$ for field states
arbitrarily close to the quantum two-mode vacuum. Moreover,
unpolarized states according to (\ref{eq:Psc}) are determined by
$\langle \op{\mathbf{S}} \rangle = \mathbf{0}$. Nonetheless, there are
states fulfilling this latter condition (as, e.g., $|n \rangle_{H} | n
\rangle_{V}$) that cannot be regarded as unpolarized, as revealed by a
number of features. These unwanted consequences call for alternative
measures.

\textit{Second-order quantum degree of polarization.---}
As from the previous discussion, it seems clear that higher-order
moments must be taken into account, as advocated by
Klyshko~\cite{Klyshko:1997}. For the time being, we concentrate on the
second order: our task is thus to link the resulting fluctuations with
our notion of a polarization degree.  To this end, we observe that a
sensible modification of (\ref{eq:Psc}) is easily obtained by
replacing $\langle \op{S}_{0} \rangle$ with $[\langle \op{S}_{0} (
\op{S}_{0} + 2 ) \rangle ]^{1/2} = \langle \op{\mathbf{S}}^{2}
\rangle^{1/2}$ in the denominator. The resulting
degree~\cite{Alodjants:1999uq}
\begin{equation}
  \label{P2alo}
  \mathbb{P}^{\prime}_{2} (\op{\varrho}) =  \sqrt{1 - 
    \frac{( \Delta \mathbf{S})^{2}}{\langle \op{\mathbf{S}}^{2} \rangle} } \, ,
\end{equation}
contains the desired second-order information (hence the subscript 2)
and fixes some of the above-mentioned problems. For example,
$\mathbb{P}^{\prime}_{2} < 1$ for every state $ | \Psi \rangle_{H} | 0
\rangle_{V}$, and $\mathbb{P}^{\prime}_{2} \rightarrow 0$ when $| \Psi
\rangle_{H} \rightarrow | 0 \rangle_{H}$. However, other bugs still
persist. The reason is that (\ref{P2alo}) does not properly represent
the behavior of the fluctuations in phase space. To catch these
aspects we propose to use
\begin{equation}
  \label{P2dp}
  \mathbb{P}_{2} (\op{\varrho}) = \sqrt{1-  \inf_{\mathbf{n}} 
    \frac{( \Delta S_{\mathbf{n}})^{2}}
    {\case{1}{3}\langle \op{\mathbf{S}}^{2} \rangle}} \, ,
\end{equation}
where $\op{S}_{\mathbf{n}} = \op{\mathbf{S}} \cdot \mathbf{n}$, with
$\mathbf{n}$ being a unit vector in an arbitrary direction of
spherical angles $(\theta, \phi)$. The factor $1/3$ has been
introduced for normalization.

To further appreciate this idea, we define the real symmetric $3
\times 3$ 
covariance matrix for the Stokes variables as $\Gamma_{k  \ell} = 
\frac{1}{2} \langle \{ \op{S}_{k}, \op{S}_{\ell} \} \rangle
- \langle \op{S}_{k} \rangle \langle \op{S}_{\ell} \rangle$, where 
$\{ , \}$ is the anticommutator~\cite{Barakat:1989ys}. In consequence, 
$( \Delta S_{\mathbf{n}} )^{2} = \mathbf{n}^{t} \, \bm{\Gamma} \,
\mathbf{n}$ and, since $\bm{\Gamma}$ is positive definite, the minimum
of $ ( \Delta S_{\mathbf{n}} )^{2}$ exists and it is unique. If we
incorporate the constraint $\mathbf{n}^{t} \mathbf{n} =1$ as a
Lagrange multiplier $\gamma$, this minimum is given by $\bm{\Gamma}
\mathbf{n} = \gamma \mathbf{n}$: the admissible values of $\gamma$ are
thus the eigenvalues of $\bm{\Gamma}$ and the directions minimizing 
$( \Delta S_{\mathbf{n}} )^{2} $ are the corresponding eigenvectors,
which, following the standard nomenclature in statistics, are known as
principal components of $\bm{\Gamma}$.

The covariance matrix $\bm{\Gamma}$ can be made diagonal by an
orthogonal matrix $\mathbf{R}$. In this rotated reference frame we
have that $\op{\bm{\mathcal{S}}} = \mathbf{R} \op{\mathbf{S}}$
satisfies $\op{\bm{\mathcal{S}}}^{2} = \op{\mathbf{S}}^{2}$, so that
\begin{equation}
  \label{eq:Jn}
  (\Delta \mathcal{S}_{1})^{2} + (\Delta \mathcal{S}_{2})^{2}  +
  (\Delta \mathcal{S}_{3})^{2} \le  \langle 
  \op{\bm{\mathcal{S}}}^{2} \rangle =  
  \langle \op{\mathbf{S}}^{2} \rangle \, ,
\end{equation}
where the subscripts 1, 2, and 3 indicate the directions of the
orthogonal eigenvectors of $\bm{\Gamma}$. The contour surface 
of these variances defines an ellipsoid that provides an accurate
representation of the noise distribution of the state.

\textit{Properties and examples.---}
Let us explore some properties of the degree $\mathbb{P}_{2}$.
Unpolarized states according to $\mathbb{P}_{2}$ are those whose
fluctuations are isotropic and saturate the bound in
Eq.~(\ref{eq:Jn}). This means that the ellipsoid reduces to a sphere
of a radius $ (\case{1}{3} \langle \op{\mathbf{S}}^{2} \rangle
)^{1/2}$.  As we shall see, this allows to distinguish hidden
polarization not revealed by $\mathbb{P}_{1}$. We note, in passing,
that the unpolarized states introduced in Ref.~\cite{Prakash:1971fr}
as those invariant under SU(2) transformations are also unpolarized
for $\mathbb{P}_{2}$. However, the converse is not true in general.

It follows directly from the definition of $\mathbb{P}_{2}$ that any
SU(2) polarization transformation $\op{U}$ leaves the degree of
polarization invariant: $\mathbb{P}_{2} (\op{\varrho} ) =
\mathbb{P}_{2} (\op{U} \, \op{\varrho} \, \op{U}^\dagger)$.

It is clear that the moments of any energy-preserving observable (such
as $\op{\mathbf{S}}$) do not depend on the coherences between
different subspaces.  This means that the only accessible information
from any state $\op{\varrho}$ is just its polarization sector, which
is defined by the block-diagonal form $\op{\varrho}_{\mathrm{pol}} =
\sum_{N=0}^\infty \op{\openone}_{N} \, \op{\varrho} \,
\op{\openone}_{N}$, where $\op{\openone}_{N}$ is the projector onto
the $N$-photon subspace. Therefore, any $\op{\varrho}$ and its
associated block-diagonal form $\op{\varrho}_{\mathrm{pol}}$ have the
same degree of polarization $\mathbb{P}_{2}$. This is consistent with
the fact that polarization and intensity are, in principle,
independent concepts: in classical optics the form of the ellipse
described by the electric field (polarization) does not depend on its
size (intensity).  All this confirms that our proposal fulfills all
the requirements for a \textit{bona fide} second-order degree of
polarization.

We further develop these ideas by presenting a few relevant
examples. First, for any two-mode number state $| n \rangle_{H} 
| m \rangle_{V}$, $\mathbb{P}_{2} ( | n \rangle_{H} | m \rangle_{V}) =
1$.  In particular, this means that $\mathbb{P}_{2}$ identifies the
hidden polarization of, e.g., the state $| n \rangle_{H} | n
\rangle_{V}$.

For two-mode quadrature coherent states $| \alpha \rangle_{H} | \beta
\rangle_{V}$, with an average number of photons $\bar{N} = | \alpha
|^2 + | \beta |^2$, simple calculations give $\mathbb{P}_{2} 
( | \alpha \rangle_{H} | \beta \rangle_{V} ) = [\bar{N}/(\bar{N} + 3)]^{1/2}$,
so when $\bar{N} \rightarrow \infty$, $\mathbb{P}_{2}$ tends
to unity, and when   $\bar{N} \rightarrow 0$, $\mathbb{P}_{2}$ tends
to zero, showing a good classical limit. Interestingly, the
covariance for these states is isotropic and the corresponding
ellipsoid reduces to a sphere of radius $\bar{N}^{1/2}$.

For single-mode states $| \Psi \rangle_{H} |0 \rangle_{V}$, we find
\begin{equation}
  \label{eq:sqv}
  \mathbb{P}_{2} ( | \Psi \rangle_{H} |0 \rangle_{V} ) = 
  \sqrt{1 - \frac{3 \min[(\Delta N)^{2}, \bar{N}]}{ ( \Delta N)^{2} + 
      \bar{N} (\bar{N} + 2 ) }} \, ,
\end{equation}
where $\bar{N}$ is the average number of photons. The problems arising
with these states when using $\mathbb{P}_{1}$ are thus avoided.

We finally consider SU(2) coherent states $| \theta, \phi \rangle =
\op{D} (\theta ,\phi ) |N, k=0 \rangle$, where $\op{D} (\theta ,\phi )
= \exp [ \theta / 2 ( \op{S}_{+} e^{-i\phi} - \op{S}_{-} e^{i\phi
})]$, with $\op{S}_{\pm} = \op{S}_{x} \pm i \op{S}_{y}$, is the
standard displacement operator on the sphere.  These are the only ones
that satisfy relation (\ref{ur}) as an equality, so $\mathbb{P}_{2} (
|\theta, \phi \rangle) = 1$ and they are completely polarized for our
approach.  Incidentally, we have also $\mathbb{P}_{1} ( |\theta, 
\phi \rangle) = 1$ and one could expect that they would fulfill similar
properties for arbitrary orders.

%%%%%%%%%%%%%%%%%%%%%%%%%%%%%%%%%%%%%%%%%%%%%%%%%%%%%%%%%%
\begin{figure}
  \includegraphics[width=0.80\columnwidth]{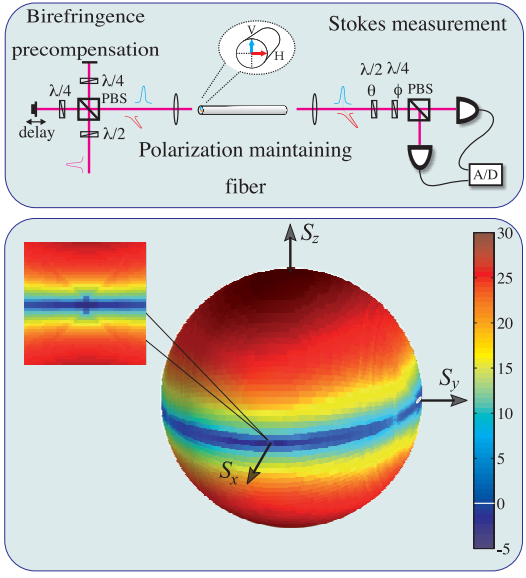}
  \caption{(Color online) (Top) Setup for efficient generation of a
    polarization squeezed state and the corresponding Stokes
    measurement apparatus.  (Bottom) Measured variances for that state
    with the indicated scale (in dB noise power relative to the shot
    noise, marked by a white line). The minimum measured variance is
    $-5.0 \pm 0.3$ dB. The white point is the tip of $\langle
    \op{\mathbf{S}} \rangle$. We include also a zoom around the
    $S_{x}$ axis, near the minimum variance.}
\end{figure}
%%%%%%%%%%%%%%%%%%%%%%%%%%%%%%%%%%%%%%%%%%%%%%%%%%%%%%%%%%%

\textit{Experiment.---} 
We demonstrate our proposal with two different quantum states of
light: a very bright polarization squeezed state and a quadrature
squeezed vacuum produced in an optical fiber and in an optical
parametric oscillator (OPO), respectively.

To generate the bright squeezed light, we employ ultrashort laser
pulses in the soliton regime of an optical fiber to achieve a large
effective nonlinear Kerr response and avoid dispersive pulse
broadening (see Fig.~1)~\cite{heersink:2005ul}.  Our experiment uses a
Cr$^{4+}$:YAG laser emitting near Fourier-limited 140~fs FWHM pulses
at 1497~nm with a repetition rate of 163~MHz. We utilize the two
polarization axes of a 13.2~m birefringent fiber to simultaneously
generate two independent quadrature squeezed states in the $H$ and $V$
modes, with a relative phase of $\pi/2$. The average output power from
the fiber was 13 mW, which, with the bandwidth definition of our
quantum state, corresponds to an average number of photons of
$10^{11}$ per 1$\mu$s.

The Stokes measurement is also shown in Fig.~1 and consists of a
half-wave plate $(\lambda /2 ,\theta)$ followed by a quarter-wave
plate ($\lambda /4, \phi$) and a polarizing beam splitter (PBS). The
transformation performed by the wave plates can be represented by
$\op{D} (\theta, \phi)$, while the PBS projects on the basis $| N, k
\rangle$. The outputs of the PBS are measured using high efficiency
photodiodes (98\%), the photocurrent difference is produced, and the
resulting fluctuations are evaluated at a sideband of 17.5~MHz (and a
bandwidth of 1~MHz). In this way, the setup enables the measurement of
$\op{S}_{\mathbf{n}}$~\cite{Marquardt:2007bh}.

For each pair of angles $(\theta, \phi)$ the noise statistics were
acquired and the optical intensities incident at both detectors were
recorded.  From this, the Stokes variances $( \Delta
S_{\mathbf{n}})^{2}$ were obtained and the results are plotted in
Fig.~1 as a color map on the sphere. The mean value $\langle
\op{\mathbf{S}} \rangle$ is parallel to the $S_{y}$ axis.

%%%%%%%%%%%%%%%%%%%%%%%%%%%%%%%%%%%%%%%%%%%%%%%%%%%%%%%%%%%%%
\begin{figure}
  \includegraphics[width=0.80\columnwidth]{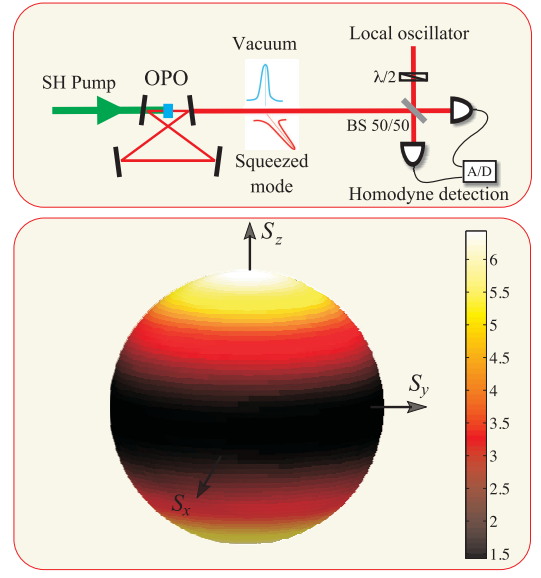}
  \caption{(Color online) (Top) Setup for the generation of a squeezed
    vacuum in the horizontal mode and vacuum in the vertical one.  We
    measure -3.8 dB (8.6 dB) of quadrature squeezing (antisqueezing)
    in the horizontal mode.  (Bottom) Measured variances $( \Delta
    S_{\mathbf{n}})^{2}$ for the state as a color map on the
    Poincar\'e sphere with the indicated (linear) scale.}
\end{figure}
%%%%%%%%%%%%%%%%%%%%%%%%%%%%%%%%%%%%%%%%%%%%%%%%%%%%%%%%%%%%%

The minimum-variance determination prescribed by $\mathbb{P}_{2}$ is
also an optimal strategy for polarization-squeezing
detection~\cite{Bowen:2002}.  In this case, it suffices to consider a
general Stokes parameter rotated by $\theta$ in the dark plane
(orthogonal to the direction of $\langle
\op{\mathbf{S}} \rangle$), namely, $\op{S}_{\theta} = \cos \theta \,
\op{S}_{x} + \sin \theta \, \op{S}_{z}$, so that $\langle
\op{S}_{\theta} \rangle = 0$.  In addition, since for bright fields
the fluctuations are small compared with the mean values, one has
$(\Delta S_{\theta})^{2} \simeq \case{1}{2} \bar{N} [ ( \Delta
X_{H,\theta})^{2} + ( \Delta X_{V,\theta})^{2} ]$, where
$\op{X}_{\theta}$ are the rotated quadratures for each polarization
mode. The searched point is obtained by optimizing over $\theta$,
finding $2^{\circ} \pm 0.3^{\circ}$~\cite{angle}.

From the data we get $\mathbb{P}_{1 } = 1$ and $\mathbb{P}_{2 } \simeq
1$ (within the experimental precision).  This is simply due to the
large excitation of the Stokes vector, which dominates in the
definition (\ref{P2dp}). In the next experiment, we produce a state
with a very small excitation, so that the second-order degree is
governed by the Stokes fluctuations.

We generate a quadrature squeezed vacuum in a well-defined
spatio-temporal polarization mode using an OPO operating below
threshold and pumped with a 532~nm light beam (see
Fig~2)~\cite{OPO}. The parametric down-conversion interaction is based
on a type I phase-matched periodically poled KTP crystal, which
produces a squeezed vacuum in the $H$ mode while leaving the $V$ mode
in the vacuum. With this information at hand, we write the resulting
state as $\varrho_H \otimes|0 \rangle_V \, {}_{V}\langle 0|$ where
$\varrho_H$ is the density operator of the state produced in the OPO.

In contrast to the previous experiment, now we characterize the
polarization by using two-mode homodyne detection. As this provides
complete knowledge about the measured state, the Stokes fluctuations
will be contained in the homodyne data. Since the $H$ and $V$
modes are known to be uncorrelated, a complete reconstruction can be
obtained just by measuring the two orthogonal modes independently. To
this end, we direct our two-mode state to a standard homodyne detector
where the polarization of the local oscillator (LO) could be swapped
between $H$ and $V$ polarizations. The measurements results are
demodulated at a sideband frequency of 5 MHz with a bandwidth of 100
kHz. The total detection efficiency is about 87~\%.

Using the time-resolved data of the orthogonal modes as well as the
\textit{a priori} state information, we fully reconstruct the density
matrix in a 16-dimensional Fock space using a maximum likelihood
algorithm. From this density matrix we calculate the moments of the
Stokes parameters $\op{S}_{\mathbf{n}}$ and plot the result as a color
map on the Poincar\'e sphere, as shown in Fig~2.  Now $\bar{N} \simeq
1.5$ and the degree of polarization of the quadrature squeezed vacuum
state is not only governed by the first moment (as the bright squeezed
state).  This is nicely illustrated in the new definition of the
degree of polarization: we calculate $\mathbb{P}_1=0.998\pm 0.001$ and
$\mathbb{P}_2=0.79\pm 0.01$. We also note that
$\mathbb{P}_2^{\prime}=0.43\pm 0.01$ and it does not capture the
variations on the Poincar\'e sphere.

\textit{Discussion and concluding remarks.---}
The definition (\ref{P2dp}) has proved to be a satisfactory solution
to deal with second-order polarization properties. Of course, a
complete characterization must involve a whole hierarchy of
polarization degrees $\mathbb{P}_{k}$ containing all the orders, as it
happens with field correlations in coherence theory. Although the
second order considered here surely accounts for most of the
interesting, and in many cases dominant effects, some subtleties may
arise when taking into account higher orders. A full analysis of these
questions exceeds the scope of this work and will be presented
elsewhere.

Our approach is based on the underlying SU(2) symmetry of light
polarization.  This makes possible a direct translation of our results
to other fields where the same symmetry plays an important role, such
as cold atoms~\cite{cold}. The approach is also well suited for other
unitary symmetries, such as SU(2)$^{\otimes n}$ or SU(3). The former
is connected with the polarization of spatial-multimode
fields~\cite{Klyshko:1992wd}, while the latter has recently attracted
a lot of attention in relation with near-field optics~\cite{near}.  In
these  cases, the optimization process can be more involved, but
the spirit of our approach remains the same.

We thank H. de Guise  for useful discussions. Financial support 
from  CONACyT (Grant  106525), the Swedish Research Council (VR) 
and the Swedish Foundation for International Cooperation in Research 
and Higher Education (STINT), DGI (Grant FIS2008-04356) and
EU project COMPAS is gratefully acknowledged.

%\bibliography{polarization}

\end{document}